\documentclass[aps,prb,preprint,groupedaddress]{revtex4}
\usepackage{amsmath}
\usepackage{amssymb}
\usepackage{graphicx}
\usepackage{dcolumn}
\usepackage{bm}

\begin{document}
\title{Supporting Material for: Superionicity and Polymorphism in Calcium 
       Fluoride at High Pressure}

\author{Claudio Cazorla}
\email{ccazorla@icmab.es}
\thanks{Corresponding Author}
\affiliation{Institute of Materials Science of Barcelona (ICMAB-CSIC),
             Campus UAB, 08193 Bellaterra, Spain} 

\author{Daniel Errandonea}
\affiliation{Departamento de F\'isica Aplicada (ICMUV),
             Universitat de Valencia, 46100 Burjassot, Spain}

\begin{abstract}
In this supplemental document we explain the details of our \emph{ab initio} density 
functional theory calculations and crystal structure searches performed in CaF$_{2}$ under 
pressure. Also we report the computed $P$-dependence of the formation energy of Frenkel pair 
defects in the cubic fluorite phase, and the vibrational phonon spectra and structural data 
of the predicted high-$T$ monoclinic $P2_{1}/c$ phase. Calculated enthalpy energies in 
compressed SrF$_{2}$ and BaF$_{2}$ are also presented. 
\end{abstract}

\maketitle

\section{Zero-temperature DFT calculations}
\label{sec:static}

\begin{figure}
\centerline{
\includegraphics[width=0.80\linewidth]{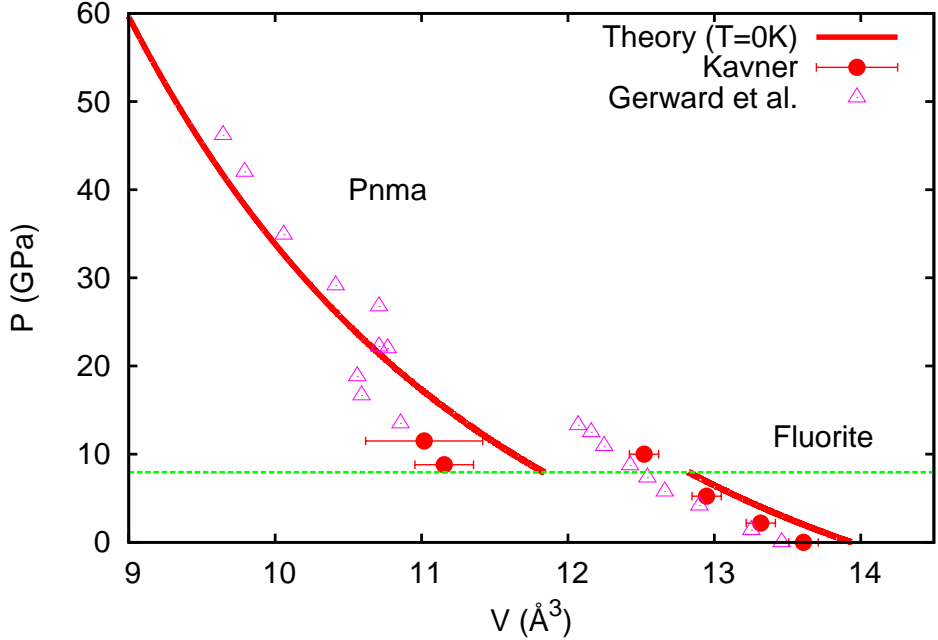}}
\caption{[Supplemental](Color online)~Calculated (solid red line) 
         and measured equation of state of CaF$_{2}$ at low temperatures. 
         Experimental data are from works~[\onlinecite{kavner08,gerward92}].}
\label{fig1}
\end{figure}

\begin{figure}
\centerline{
\includegraphics[width=0.80\linewidth]{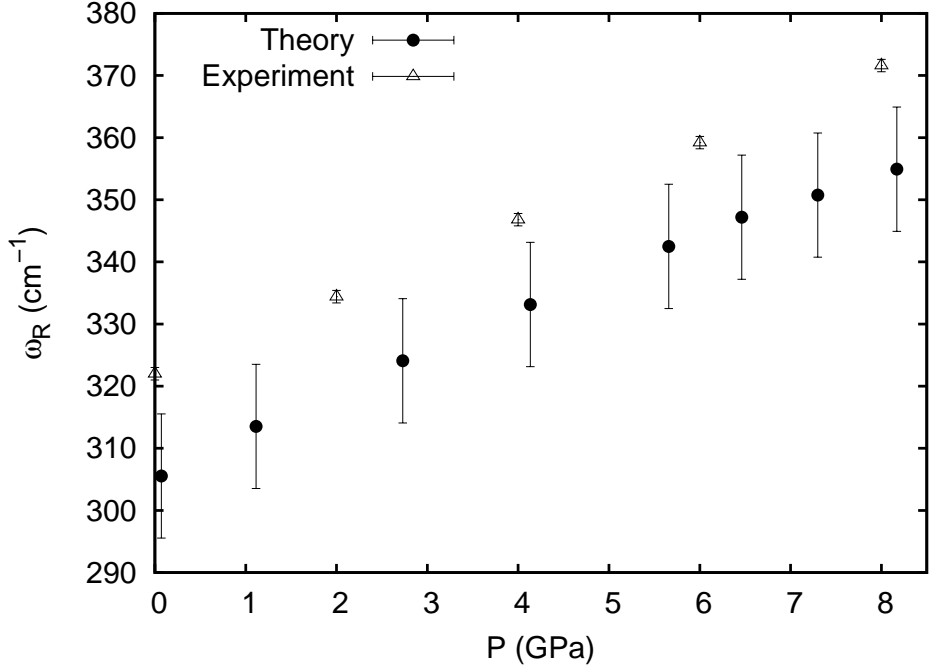}}
\caption{[Supplemental]~Measured and calculated Raman frequencies
         in $\alpha$-CaF$_{2}$ at low temperatures and expressed as a 
         function of pressure.}
\label{fig2}
\end{figure}

Total energy calculations were carried out in CaF$_{2}$ under pressure in the cubic fluorite 
and orthorhombic cotunnite phases, employing the methods and parameters described in 
the main text of the present Letter. As it may be appreciated in the supplemental Fig.~\ref{fig1}, 
the agreement between our calculated equation of state and experimental data from 
Refs.~[\onlinecite{kavner08,gerward92}] is remarkably good.  

In order to calculate the vibrational phonon frequencies of CaF$_{2}$, we employed
the direct approach~[\onlinecite{kresse95,alfe01}]. In this method, the force-constant matrix is directly 
calculated in real-space by considering the proportionality between the atomic displacements 
and forces when the former are sufficiently small. Large supercells must then be constructed 
in order to guarantee that the elements of the force-constant matrix have all fallen off to 
negligible values at their boundaries, a condition that follows from the use of periodic boundary
conditions. In the present work, we employed simulation boxes larger than $216$ atoms and 
dense ${\bf k}$-point grids in the calculation of the atomic forces with VASP. 
Computation of the nonlocal parts of the pseudopotential contributions was performed in reciprocal, 
rather than real, space. Once the force-constant matrix is thus obtained, we can Fourier transform 
it to obtain the phonon spectrum at any ${\bf q}$ point. This step was done with the PHON code due 
to Alf\`e~[\onlinecite{alfe09a}]. In using this code, we exploited the translational invariance of the 
system to impose the three acoustic branches to be exactly zero at the ${\bf q}$ point, and used 
central differences in the atomic forces (i.e., we considered positive and negative atomic displacements).

We must note, however, that convergence of the force-constant matrix elements with respect to the size 
of the supercell in ionic materials may be slow due to the appearance of charge dipoles and macroscopic 
electric fields in the limit of zero wave vector. Fortunately, long-range dipole-dipole interactions can 
be modeled at the harmonic level from knowledge of the atomic Born effective charge tensors and the dielectric 
tensor of the material~[\onlinecite{gonze89,baroni01}].
Taking advantage of this result, Wang \emph{et al.} proposed a mixed-space approach in which accurate 
force constants are calculated with the direct approach in real space and long-range dipole-dipole 
interactions with linear response theory in reciprocal space~[\onlinecite{wang10,cazorla13}]. In the present
work, we used Wang \emph{et al.} mixed-space approach to calculate the vibrational phonon frequencies
of CaF$_{2}$ at different volumes.

In the supplemental Fig.~\ref{fig2}, the Raman frequencies that we have measured ($T = 300$~K) and 
calculated ($T = 0$~K) in $\alpha$-CaF$_{2}$ under compression are represented. The agreement between our 
measurements and calculations is good (i.e., within the $5$~\% of difference), especially in 
what concerns the variation of the Raman frequency with pressure which is found to be almost constant and
equal to $6.0$~cm$^{-1}$/GPa in the experiments and $6.2$~cm$^{-1}$/GPa in the simulations. Both our experimental and theoretical results are in good agreement with previous experiments~[\onlinecite{speziale02}].

\section{One-phase and Two-phase coexistence \emph{ab initio} molecular dynamics simulations}
\label{sec:dynamic}

\begin{figure}
\centerline{
\includegraphics[width=0.70\linewidth]{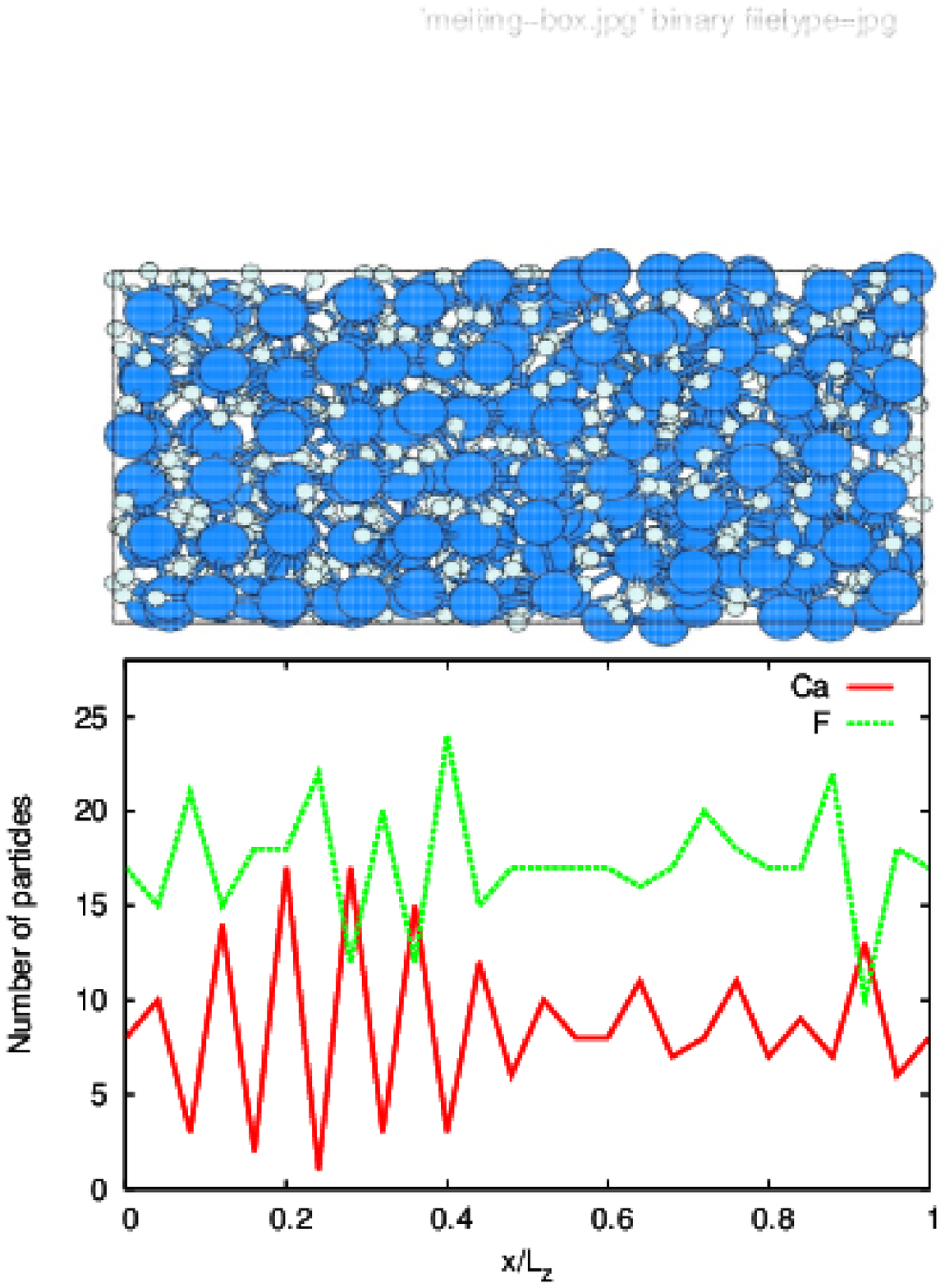}}
\caption{[Supplemental](Color online)~Determination of melting temperatures
                based on two-phase coexistence \emph{ab initio} molecular dynamics 
                simulations.
                \emph{Top}: Atomic positions projected over the $x-y$ plane of
                 the simulation box. Calcium and fluorine ions are represented
                 with large~(dark blue) and small~(bright blue) spheres, respectively.
                \emph{Bottom}: Number of particles histogram represented as a function
                of position along the direction parallel to the initial superionic-liquid
                boundary.}
\label{fig3}
\end{figure}

Our \emph{ab initio} molecular dynamics (AIMD) calculations were of two types: one-phase 
(i.e., solid and superionic phases) and two-phase coexistence (i.e., liquid and superionic 
phases coexisting in thermodynamic equilibrium) simulations.
One-phase simulations were performed in the canonical $( N, V, T )$
ensemble while two-phase coexistence simulations in the microcanonical $( N, V, E )$ ensemble.
In the $( N, V, T )$ simulations the temperature was kept fluctuating around a set-point
value by using Nose-Hoover thermostats. Large simulation boxes containing $192$
and $648$ atoms were used in our one-phase and two-phase coexistence simulations,
respectively. Periodic boundary conditions were applied along the three Cartesian directions
in all the calculations. Newton's equations of motion were integrated using the customary
Verlet's algorithm and a time-step length of $10^{-3}$~ps.
$\Gamma$-point sampling for integration within the first Brillouin zone was employed in all 
our AIMD simulations.

Comprehensive one-phase $(N , V, T)$ molecular dynamics simulations were carried out in order 
to compute the $\alpha-\beta$ and $\delta-\epsilon$  phase boundaries in CaF$_{2}$ as a function 
of pressure. Calculations comprised large simulation boxes and long simulation times of up to 
$\sim 30$~ps. We systematically carried out simulations at temperature intervals of
$250$~K, from $1000$ up to $3500$~K, at each considered volume.

Following previous works~[\onlinecite{cazorla07a,cazorla07b,cazorla09,cazorla11,cazorla12,alfe09}], 
we performed comprehensive $( N, V, E )$ two-phase coexistence MD simulations in order to determine 
the melting curve of CaF$_{2}$ at low and high pressures. Starting with a supercell containing 
the perfect crystal structure (i.e. either cubic fluorite or orthorhombic $Pnma$), we thermalize it 
at a temperature slightly below the expected melting temperature for about $4$~ps. The system remains 
in a superionic state.
The simulation is then halted and the positions of the atoms in one half of the supercell are 
held fixed while the other half is heated up to a very high temperature (typically five times 
the expected melting temperature) for about $4$~ps, so that it melts completely. With the fixed 
atoms still fixed, the molten part is rethermalized to the expected melting temperature (for about 
$2$~ps). Finally, the fixed atoms are released, thermal velocities are assigned, and the whole system
is allowed to evolve freely at constant $( N, V, E )$ for a long time (normally more than $20$~ps), 
so that the solid and liquid come into equilibrium.
The system is monitored by calculating the average number of particles in slices of the cell
taken parallel to the boundary between the solid and liquid. With this protocol, there is a
certain amount of trial and error to find the overall volume which yields the coexisting solid
and liquid system. An example of a successful coexistence run is shown in the 
supplemental Fig.~\ref{fig3}.

\section{Formation energy of Frenkel pair defects under pressure}
\label{sec:frenkel}

We calculated the formation energy of Frenkel pair defects (FPD)
in cubic CaF$_{2}$ at different pressures. We used a simulation box
containing $96$ atoms and a $2 \times 2 \times 2$ Monkhorst-Pack
${\bf k}$-point grid for integrations within the first Brillouin zone. 
We found that the series of calculated energies (see the supplemental 
Fig.~\ref{fig4}) could be very well fitted to the power law function
\begin{equation}
E_{FPD} (P) = E_{0} + a \cdot P^{b}~, 
\end{equation} 
where the optimal values of the parameters are $E_{0} = 2.07$~eV, $a = 0.054$~eV 
and $b = 0.851$. 
It is noted that the formation energy of FPD increases monotonically and 
appreciably with compression.

\begin{figure}
\centerline{
\includegraphics[width=0.80\linewidth]{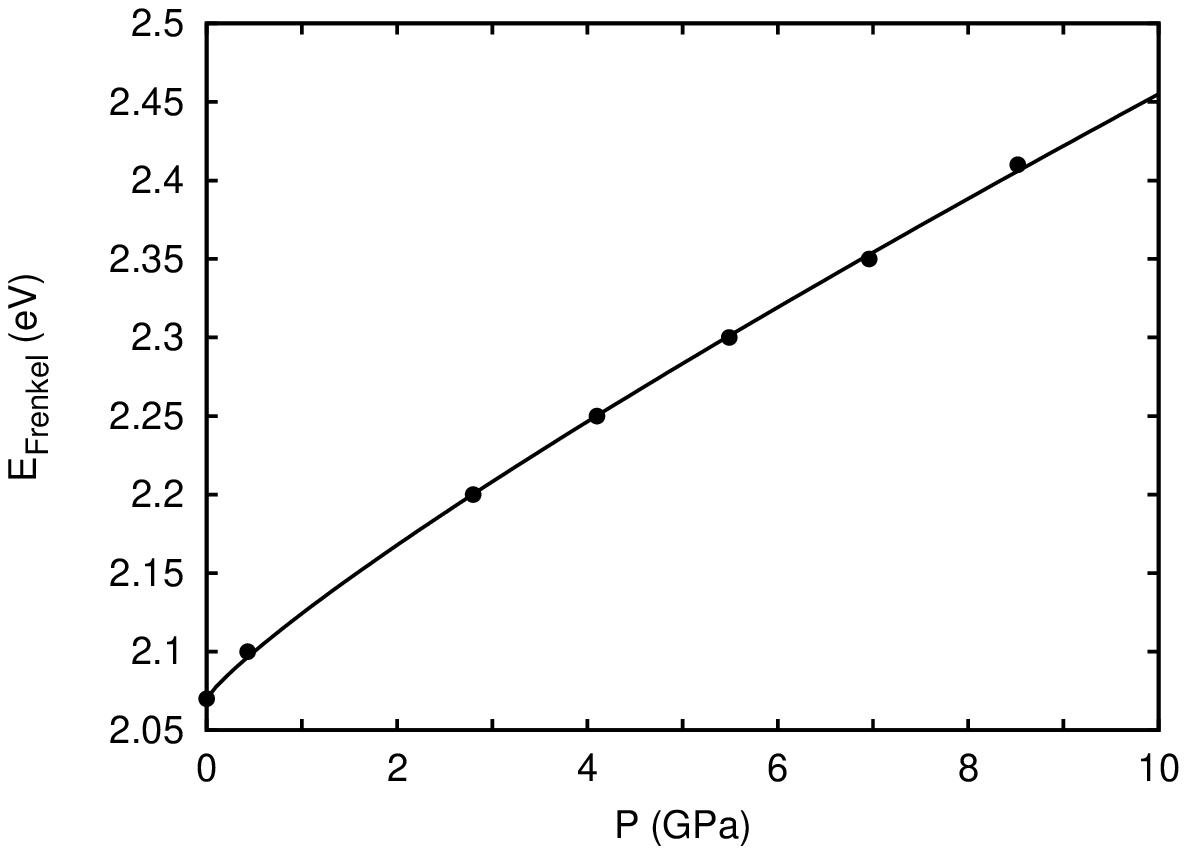}}
\caption{[Supplemental]~Calculated formation energy of Frenkel 
         pair defects in $\alpha$-CaF$_{2}$ under pressure. 
         The solid line represents a numerical fit to the
         series of computed points (see text).}
\label{fig4}
\end{figure}

\section{Crystal structure searches}
\label{sec:crystsearch}

Our crystal structure searches performed in compressed CaF$_{2}$ relied on two different
strategies, namely (i)~comprehensive atomic relaxations of configurations generated in 
constrained low-$T$ AIMD runs, and (ii)~enthalpy calculation of phases that have been recently
observed and predicted in isomorphic AB$_{2}$ compounds under pressure.

In our restricted low-$T$ AIMD runs, we used a small simulation cell containing 
$24$ atoms in the orthorhombic $Pnma$ phase and set the temperature to ambient 
conditions. Consistently to the observed $\gamma \to \delta$ transition, we constrained 
the Ca$^{2+}$ cations not to move during the dynamical simulations. The total duration of 
these AIMD runs was $\sim 20$~ps. Once these simulations were finished, we randomly selected 
$100$ configurations out of $10,000$ and performed atomic and crystal cell relaxations in them. 
During the geometry optimizations all cations and anions were allowed to move. Subsequently, 
we identified the crystal symmetry of the final equilibrium structures with the ISOTROPY 
package~[\onlinecite{isotropy}]. 
Following this strategy, we obtained the five candidate high-$T$ crystal phases that we mention 
in the main text of the present Letter (i.e., orthorhombic $P2_{1}2_{1}2_{1}$ and $Pmn2_{1}$, and
monoclinic $Pc$, $P2_{1}$ and $P2_{1}/c$).         

We also computed the energy of the post-cotunnite $P112_{1}/a$ and $P2_{1}/n$ phases in CaF$_{2}$ under 
pressure. These phases have been recently observed and predicted in bulk AuIn$_{2}$
and AuGa$_{2}$ respectively~[\onlinecite{godwal10,godwal13}], which are isomorphic materials to 
CaF$_{2}$. By following this strategy, however, we did not find any promising high-$T$ candidate crystal phase since their corresponding enthalpies were too large in comparison to those of 
the crystal structures obtained with strategy (i).

\section{Structural data, phonons and enthalpy of the predicted high-$T$ monoclinic $P2_{1}/c$ phase}
\label{sec:p21c}

\begin{figure}
\centerline{
\includegraphics[width=0.40\linewidth]{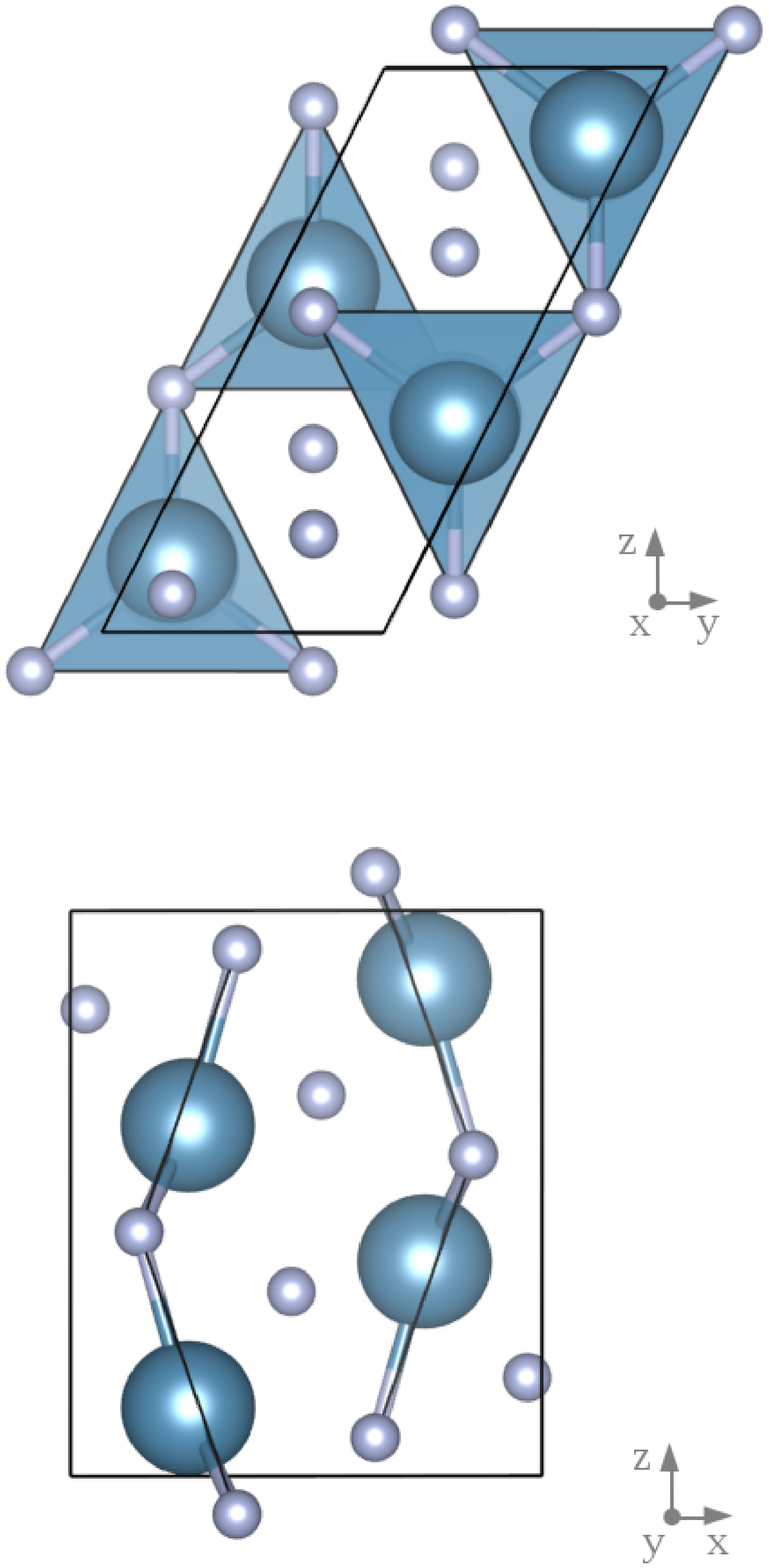}}
\caption{[Supplemental](Color online)~Sketch of the predicted 
         high-$T$ monoclinic $P2_{1}/c$ phase. Large~(Blue) 
         and small~(grey) spheres represent calcium and fluorine
         atoms, respectively.}
\label{fig5}
\end{figure}

\begin{figure}
\centerline
        {\includegraphics[width=0.80\linewidth]{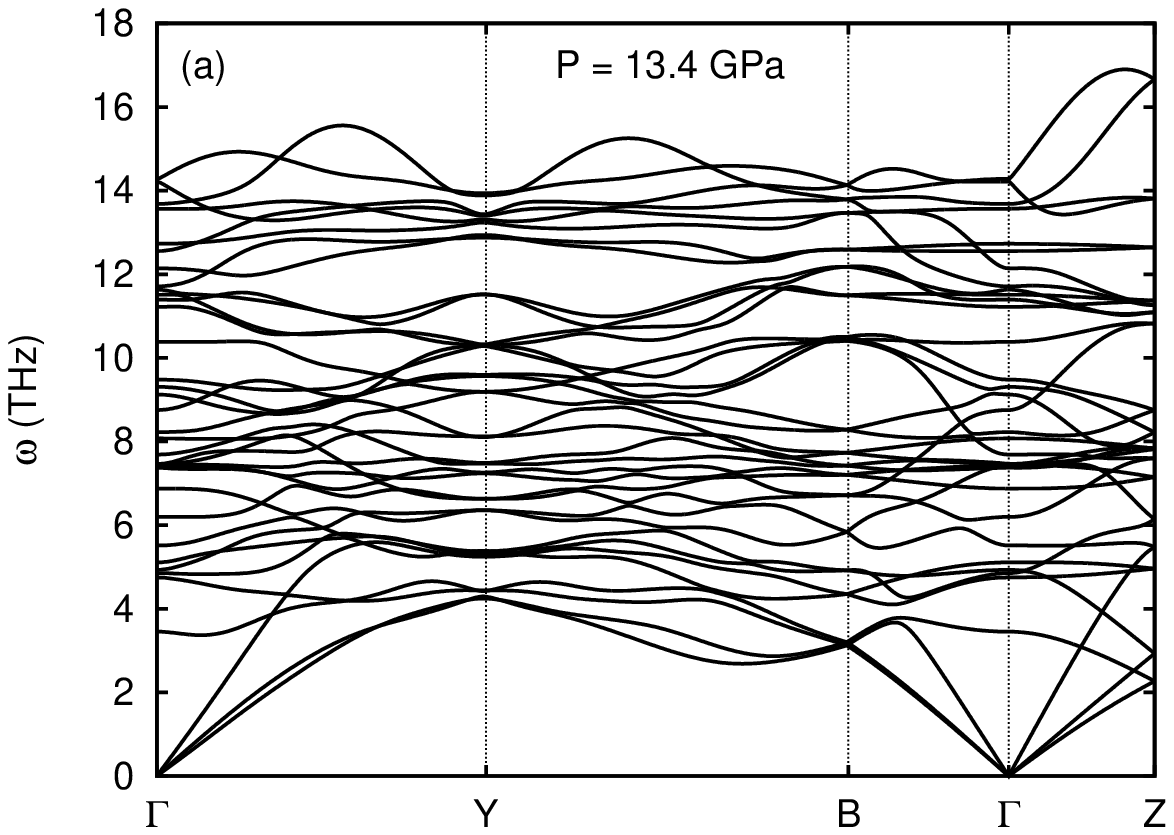}}
        {\includegraphics[width=0.80\linewidth]{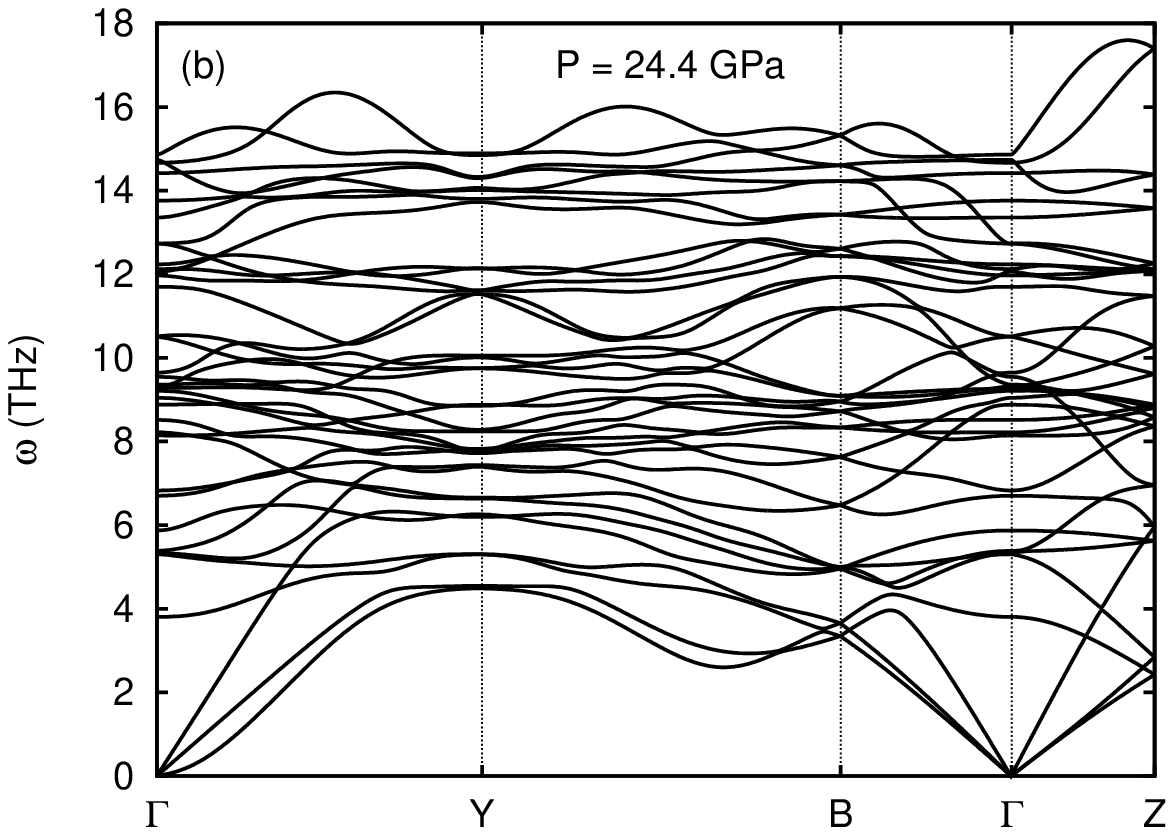}}
\caption{[Supplemental]~Calculated phonon spectrum of the predicted
         high-$T$ monoclinic $P2_{1}/c$ phase at different pressures.}
\label{fig6}
\end{figure}

\begin{table}
\begin{center}
\label{tab:structure}
\begin{tabular}{c c c c c}
\hline
\hline
$  $ & $  $ & $  $ & $  $ & $ $ \\
\multicolumn{2}{c}{$P2_{1}/c$} & $ a = 3.347$~\AA &  $ b = 6.032$~\AA  &  $ c = 7.477$~\AA  \\
\multicolumn{2}{c}{${\rm (P~=~13.4~GPa)}$} & $ \alpha = 90~^{\circ}$ &  $ \beta = 116.6~^{\circ}$  &  $ \gamma = 90~^{\circ} $  \\
$  $ & $  $ & $  $ & $  $ & $ $ \\
\hline
$  $ & $  $ & $  $ & $  $ & $ $ \\
${\rm Atom} $ & $ {\rm Wyc.} $ & $ x $ & $ y $ & $ z $ \\
$  $ & $  $ & $  $ & $  $ & $ $ \\
${\rm Ca} $ & $ 4e $ & $ 0.37156 $ & $ 0.75481 $ & $ 0.12145 $ \\
${\rm F}  $ & $ 4e $ & $ 0.31867 $ & $ 0.14034 $ & $ 0.06867 $ \\
${\rm F } $ & $ 4e $ & $ 0.08291 $ & $ 0.53005 $ & $ 0.83343 $ \\
$  $ & $  $ & $  $ & $  $ & $ $ \\
\hline
\hline
$  $ & $  $ & $  $ & $  $ & $ $ \\
\multicolumn{2}{c}{$P2_{1}/c$} & $ a = 3.353$~\AA &  $ b = 5.620$~\AA  &  $ c = 7.477$~\AA  \\
\multicolumn{2}{c}{${\rm (P~=~24.4~GPa)}$} & $ \alpha = 90~^{\circ}$ &  $ \beta = 116.6~^{\circ}$  &  $ \gamma = 90~^{\circ} $  \\
$  $ & $  $ & $  $ & $  $ & $ $ \\
\hline
$  $ & $  $ & $  $ & $  $ & $ $ \\
${\rm Atom} $ & $ {\rm Wyc.} $ & $ x $ & $  y $ & $  z $ \\
$  $ & $  $ & $  $ & $  $ & $ $ \\
${\rm Ca} $ & $ 4e $ & $ 0.36963 $ & $ 0.75135 $ & $ 0.11987 $ \\
${\rm F } $ & $ 4e $ & $ 0.31873 $ & $ 0.14546 $ & $ 0.06872 $ \\
${\rm F } $ & $ 4e $ & $ 0.07811 $ & $ 0.52958 $ & $ 0.82824 $ \\
$  $ & $  $ & $  $ & $  $ & $ $ \\
\hline
\hline
\end{tabular}
\end{center}
\caption{[Supplemental]~Structural data of the predicted high-$T$ monoclinic
         $P2_{1}/c$ phase calculated at different pressures.
         Wyckoff positions were generated with the ISOTROPY package~\cite{isotropy}.}
\end{table}

\begin{figure}
\centerline
        {\includegraphics[width=0.80\linewidth]{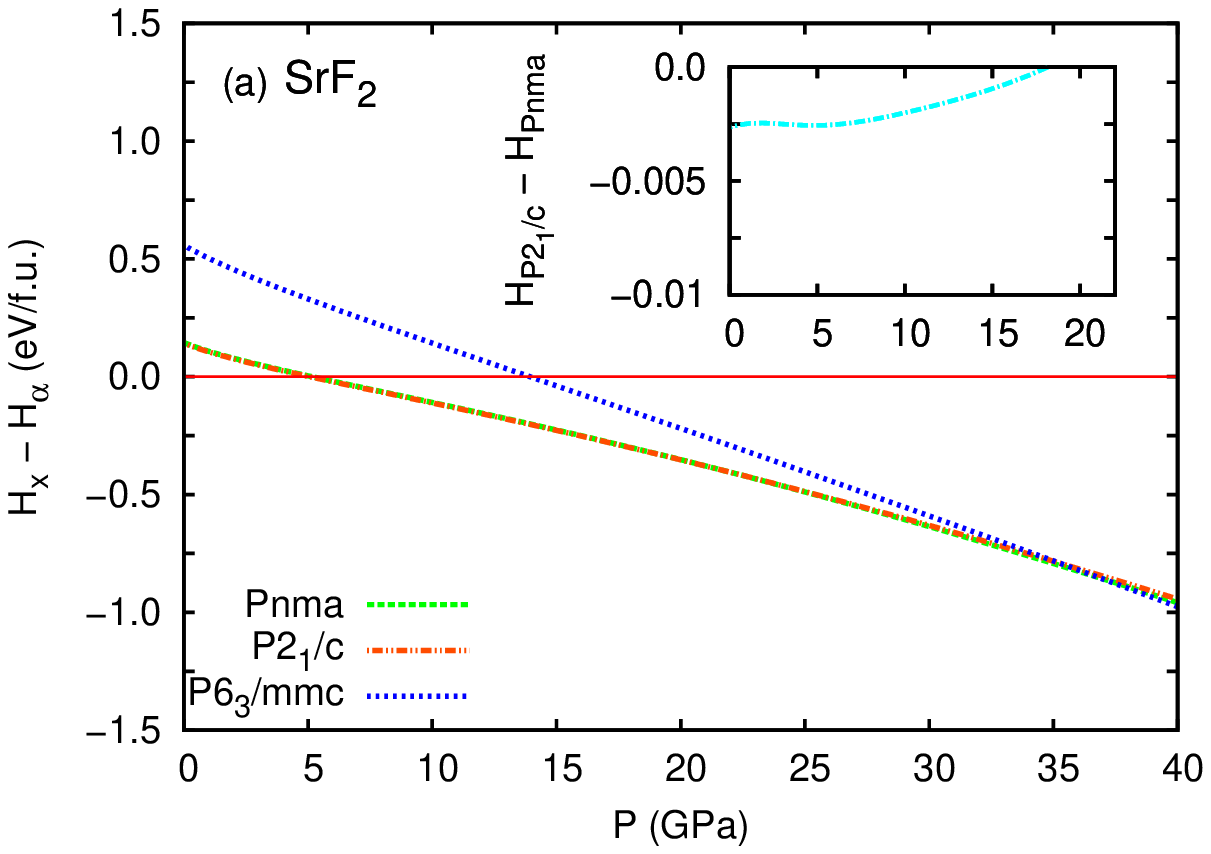}}
        {\includegraphics[width=0.80\linewidth]{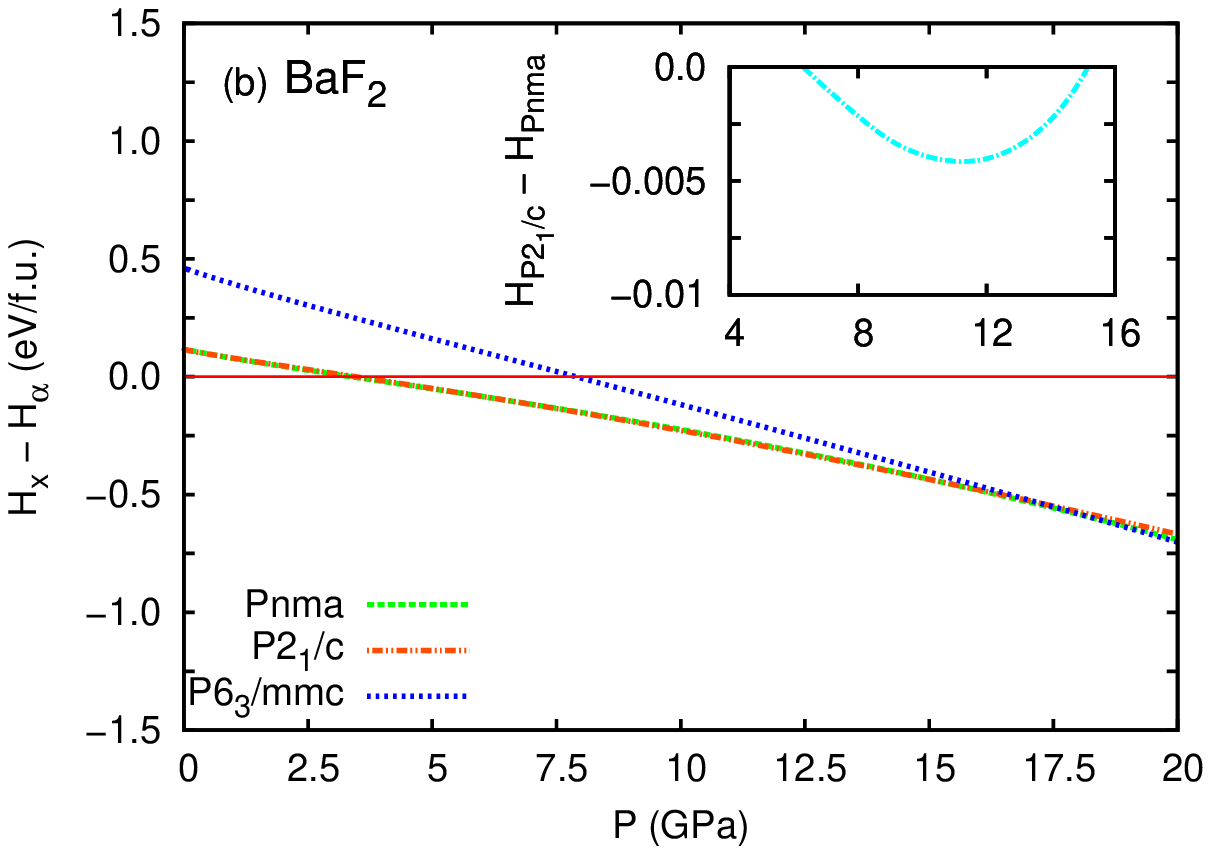}}
\caption{[Supplemental]~Calculated enthalpy of several crystal structures referred to 
         that of the cubic $\alpha$ phase and expressed as a function of pressure
         in SrF$_{2}$ and BaF$_{2}$.
         \emph{Inset}: Detail of the enthalpy difference between the $P2_{1}/c$
         and $Pnma$ phases.}
\label{fig7}
\end{figure}

A sketch of the new high-$T$ monoclinic $P2_{1}/c$ phase predicted in CaF$_{2}$ at high-$P$
is shown in the supplemental Fig.~\ref{fig5}. This structure has a similar cation coordiantion polyhedra to
that of the orthorhombic $Pnma$ phase. On it each calcium is coordinated to $9$ fluorine atoms 
that form an elongated tricapped trigonal prism. Structural data of this monoclinic phase
obtained at different pressure conditions are enclosed in Table~I.

We computed the phonon spectrum of this high-$T$ monoclinic phase at different pressures
employing the methodology explained in this supplemental document (see Sec.~\ref{sec:static}), 
and found that it was vibrationally and mechanically stable (see the supplemental Fig.~\ref{fig6}).

The results of our enthalpy calculations in SrF$_{2}$ and BaF$_{2}$ compounds
are enclosed in the supplemental Fig.~\ref{fig7}. In particular, we computed the enthalpy energy
of the cubic fluorite, orthorhombic $Pnma$, hexagonal $P6_{3}/mmc$, and
monoclinic $P2_{1}/c$ phases in the corresponding pressure regimes of interest. As we
found in CaF$_{2}$, the monoclinic $P2_{1}/c$ phase turns out to be energetically very
competitive with respect to the orthorhombic $Pnma$ phase, and actually under some
conditions the enthalpy difference between the two phases is zero within our numerical
accuracy (i.e., $3$~meV/f.u.).

\end{document}